\documentclass{PoS}

\usepackage{graphicx,psfrag,pifont,setspace,subfigure}
\usepackage{amsfonts,amsthm,bm,bbm,amsmath,amssymb,marvosym}

\title{Center vortex influence on the Dirac spectrum }

\ShortTitle{Center vortex influence on the Dirac spectrum }

\author{\speaker{Urs Heller}\\
  American Physical Society, One Research Road, Box 9000, Ridge,
  NY 11961-9000, USA\\
  E-mail: \email{heller@aps.org}}
\author{R. H\"ollwieser\\
  Atomic Institute, Technical University of Vienna, Wiedner Hauptstr.\ 8-10,
  A-1040 Vienna, Austria\\
  E-mail: \email{hroman@kph.tuwien.ac.at}}
\author{M. Faber\\
  Atomic Institute, Technical University of Vienna, Wiedner Hauptstr.\ 8-10,
  A-1040 Vienna, Austria\\
  E-mail: \email{faber@kph.tuwien.ac.at}}
\author{J. Greensite\\
  Physics and Astronomy Dept., San Francisco State University, San Francisco,
  CA~94132, USA\\
  E-mail: \email{jgreensite@gmail.com}}
\author{{\v S}. Olejn\'{\i}k\\
  Institute of Physics, Slovak Academy of Sciences, SK--845 11 Bratislava,
  Slovakia\\
  E-mail: \email{stefan.olejnik@gmail.com}}

\abstract{We study the influence of center vortices on the low-lying
eigenmodes of the Dirac operator, in both the overlap and asqtad
formulations.   For center-projected configurations, one finds that
the low-lying near-zero modes are present in the staggered (asqtad)
formulation, but not in the overlap and ``chirally-improved"
formulations. We argue that this is due to the absence of a
field-independent chiral symmetry in the latter formulations, when the
Dirac operator is evaluated on the very rough configurations generated by
center projection. We also confirm and extend the results of Kovalenko
et al.\ [Phys. Lett. B 648, 383 (2007)], finding strong correlations
between center vortex locations, and the scalar density of low-lying
Dirac eigenmodes on unprojected lattices, in both asqtad and overlap
formulations. It is found that the low-lying eigenmodes have their largest
concentrations in point-like regions, rather than on submanifolds of
higher dimensionality.}

\FullConference{The XXVI International Symposium on Lattice Field Theory \\
		 July 14 - 19, 2008\\
		 Williamsburg, Virginia, USA}

\newcommand{\beq}{\begin{equation}}
\newcommand{\eeq}{\end{equation}}
\newcommand{\bea}{\begin{eqnarray}}
\newcommand{\eea}{\end{eqnarray}}
\newcommand{\cb}{$\chi SB$}

\renewcommand{\l}{\lambda}

\renewcommand{\b}{\beta}

\newcommand{\m}{\mu}

\begin{document}

\section{Introduction}\label{Intro}

Center vortices were introduced to explain quark confinement, and there
are good reasons to believe that a force strong enough to confine
quarks must also break chiral symmetry spontaneously \cite{Casher}.
Several years ago, however, Gattnar et al.\ \cite{Gattnar} reported
a puzzling result concerning the low-lying eigenvalue spectrum of a
chirally-improved version of the Dirac operator \cite{Gattringer},
when evaluated on center projected lattices.  Despite the confining
properties of such lattices, these authors found a large gap in the
spectrum around zero eigenvalue.  This implies, via the Banks-Casher
relation \cite{BC}, a vanishing chiral condensate, and unbroken chiral
symmetry.  In the present work we suggest that the large gap found in
the chirally-improved and overlap Dirac operators, when evaluated on
confining, center-projected configurations, is related to the way in
which chiral symmetry is realized on the lattice. The Casher argument
\cite{Casher} for chiral symmetry breaking (\cb) is based on the usual
$SU(N_f)_L \times SU(N_f)_R$ symmetry of the continuum theory with
massless fermions, with symmetry transformations that are independent
of the gauge-field configuration. For overlap fermions, however, the
chiral symmetry transformations are gauge-field independent only for
configurations which vary smoothly at the lattice scale, while for the
chirally-improved Dirac operator due to Gattringer,  chiral symmetry
itself is absent for non-smooth configurations.  Center-projected
configurations are as far from smooth as possible.  Thus, for the overlap
and chirally-improved Dirac operators evaluated on these configurations,
the direct connection to continuum chiral symmetry is lost, and the
Casher argument for spontaneous chiral symmetry breaking need not apply.

In section 2 we display the spectra of the overlap~\cite{overlap}
and asqtad~\cite{ASQTAD} Dirac operators, when evaluated on normal,
vortex-only (i.e.\ center-projected), and vortex-removed lattices. The
asqtad operator has a field-independent remnant chiral symmetry,
and the eigenvalue gap disappears.  Our results support the view that
center vortices alone can induce both confinement \emph{and}  chiral
symmetry breaking.\footnote{For related results, cf.\  Alexandrou et
al.\ \cite{AdFE}, Gubarev et al.\ \cite{morozov}, and Bornyakov et al.\
\cite{Borny}.} In section 3 we report on other correlations between
center vortex location, and the density distribution of low-lying
Dirac eigenmodes, following the earlier work by Kovalenko et al.\
\cite{ITEP}. These correlations are consistent with the picture
advocated by Engelhardt and Reinhardt \cite{ER}, in which topological
charge is concentrated at points where vortices either intersect,
or twist about themselves (``writhe'') in a certain way. Dirac zero
modes are concentrated where the topological charge density is large,
and therefore one would expect that the densities of low-lying eigenmodes
would be peaked in point-like regions. We provide supporting evidence for
this type of concentration.   We work throughout with lattices generated
by lattice Monte Carlo simulation of the tadpole improved L\"uscher-Weisz
pure-gauge action~\cite{Luscher:1985zq}, mainly at coupling $\b_{LW}=3.3$
(lattice spacing $a=0.15$ fm) for the $SU(2)$ gauge group. Center
projection is carried out after fixing to the direct maximal center gauge.

\section{Thin Vortices and Near-Zero Modes}\label{sec:vortices}

\begin{figure*}[h]
  \centering
  \psfrag{ReL}{Re $\lambda$}
  \psfrag{ImL}{Im $\lambda$}
  \psfrag{p}{\footnotesize $\;\;\quad$periodic}
  \psfrag{a}{\footnotesize $\qquad$antiperiodic}
  \psfrag{b}{\footnotesize boundary conditions}
  \includegraphics[scale=0.75]{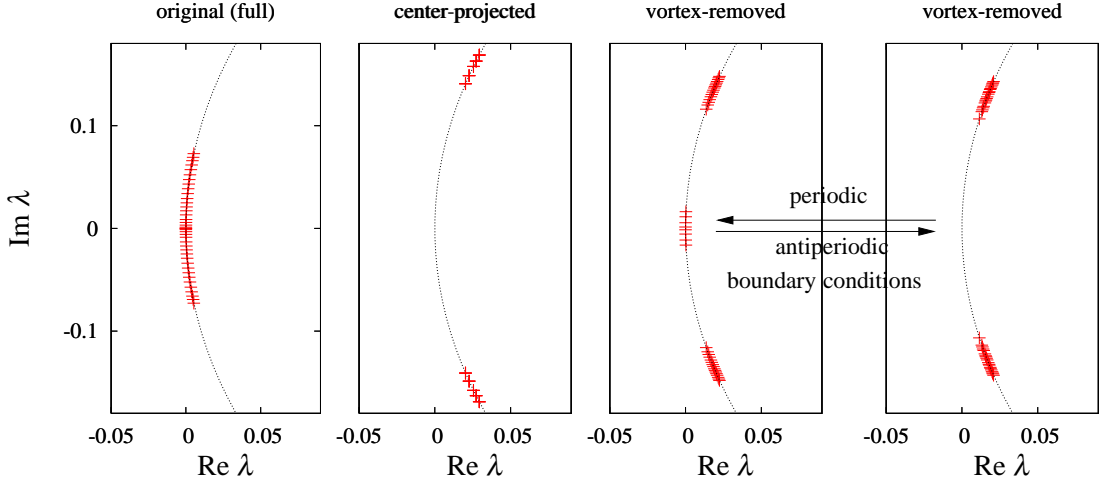}
  \caption{The first twenty overlap Dirac eigenvalue pairs on the
  Ginsparg-Wilson circle for a $16^4$ lattice at $\b_{LW}=3.3$. The
  center-projected configurations show a four-fold degeneracy. Zero-modes
  in vortex-removed configurations disappear for antiperiodic boundary
  conditions.}
\label{fig:ovlevs}
\end{figure*}

\begin{figure*}[t]
  \centering
  \psfrag{ReL}{Re $\lambda$}
  \psfrag{ImL}{Im $\lambda$}
  \includegraphics[scale=0.7]{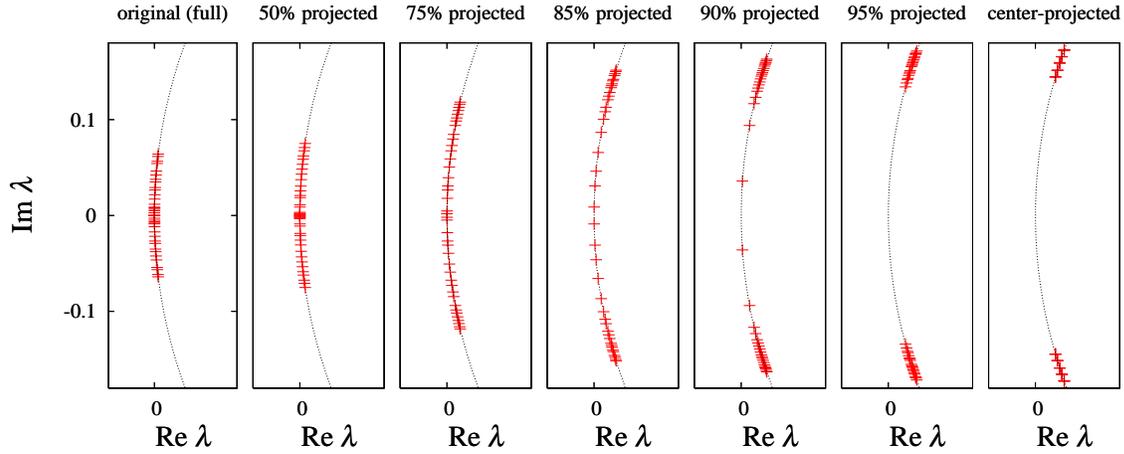}
  \caption{The first twenty overlap Dirac eigenvalue pairs from a single
  configuration on a $16^4$ lattice, antiperiodic boundary conditions at
  $\b_{LW}=3.3$, for interpolated fields.}
  \label{fig:interevs}
\end{figure*}

Fig.~\ref{fig:ovlevs} displays the first twenty overlap eigenvalue pairs
for a $16^4$ lattice at $\beta_{LW}=3.3$. There is a large gap around zero
for the center-projected data, which implies zero chiral condensate. In
this case we see only five distinct eigenvalue pairs.  This is due to the
fact that in center projection with $U_\mu(x) = \pm\mathbbm 1_2$, the two
colors decouple and the eigenvalue equation $D \psi_n = \l_n \psi_n$ is
invariant under charge conjugation.\footnote{This assumes that the Dirac
operator has the Wilson or overlap (but not staggered) form. Thus, if
$\psi_n$ is an eigenstate with eigenvalue $\l_n$, then $C^{-1} \psi_n^*$
is also an eigenstate, with the same eigenvalue \cite{Leutwyler:1992yt}.}
A four-fold degeneracy results.  The vortex-removed data shows four
near-zero modes for each chirality.  These correspond in number to the
zero modes of the free theory, independent of lattice size, and disappear
when antiperiodic boundary conditions are imposed.  They are irrelevant to
\cb. We have argued that the reason for the large gap in the vortex-only
case is the lack of smoothness of center-projected lattices, which results
in a strong field-dependence of the chiral symmetry of the Dirac operator,
in contrast to the symmetry of the continuum theory.   If this is indeed
the reason for the gap, then the gap should disappear, and \cb~ should be
recovered, under a suitable smoothing of the center-projected lattice.
We therefore perform an interpolation between full and projected
configurations, reducing separation in the group manifold between
each link variable $U_\m(x)$ in maximal center gauge, and its nearest
center element, by some fixed percentage. In Fig.\  \ref{fig:interevs}
we show the low-lying eigenvalues for partial projections together with
the unprojected and center-projected lattices.  We see that there is no
really obvious gap in the partially-projected lattices, even at $85\%$
projection. This agrees with our conjecture that applying the overlap
operator to a smoother version of the vortex-only vacuum would give a
result consistent with \cb ~and the Banks-Casher relation.
Staggered and asqtad fermions, on the other hand, do not require a smooth
configuration to preserve a subgroup of the usual continuum $SU(N_f)_L
\times SU(N_f)_R$ symmetry, and by the Casher argument \cite{Casher}
one would expect this remaining symmetry to be spontaneously broken by
any confining gauge configuration. Indeed, Ref.\ \cite{AdFE} already
reported that $\langle \overline{\psi} \psi \rangle > 0$ for staggered
fermions on a center-projected lattice.

\begin{figure*}[h]
  \centering
  \psfrag{ReL}{Re $\lambda$}
  \psfrag{ImL}{Im $\lambda$}
  \psfrag{p}{\footnotesize $\;\;\quad$periodic}
  \psfrag{a}{\footnotesize $\qquad$antiperiodic}
  \psfrag{b}{\footnotesize boundary conditions}
  \includegraphics[scale=0.7]{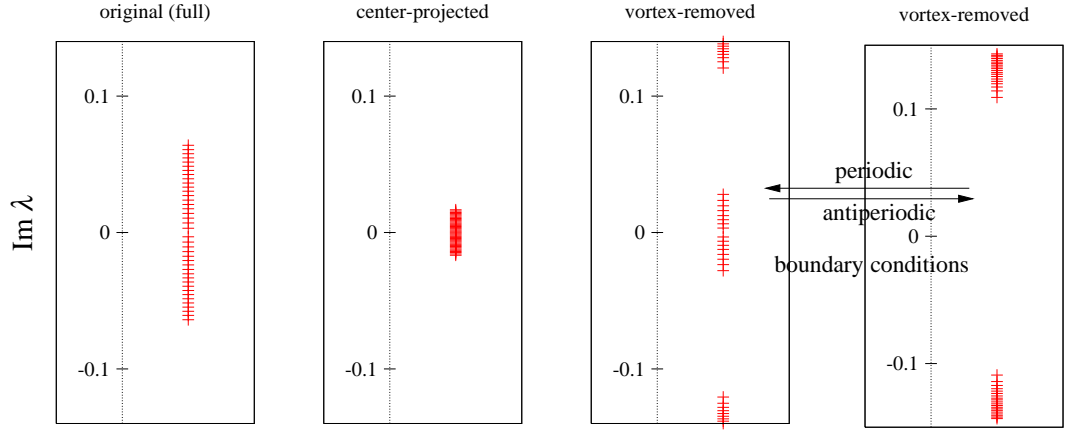}
  \caption{The first twenty asqtad Dirac eigenvalue pairs from a $16^4$
  lattice at $\b_{LW}=3.3$. The center-projected configurations show no
  gap around zero. Zero-modes in vortex-removed configurations disappear
  for antiperiodic boundary conditions.}
\label{fig:stevs}
\end{figure*}

Fig.~\ref{fig:stevs} shows the first twenty asqtad eigenvalue pairs. The
low-lying eigenmode density actually increases for center-projected
compared to unmodified lattices; the gap found in the overlap and
chirally-improved formulations has disappeared. Thus, for the asqtad
operator, we have found exactly what was expected prior to the results
of Gattnar et al.\ \cite{Gattnar}:  the vortex excitations of the
vortex-only lattice carry not only the information about confinement,
but are also responsible for \cb~ via the Banks-Casher relation. The
vortex-removed data develops a central band around Im$\lambda=0$ of eight
doubly degenerate eigenmodes per chirality, which are a remnant of the
32 free-field zero modes (four zero modes for each of four ``tastes''
times two colors), and play no role in \cb. These modes again disappear
using antiperiodic boundary conditions in one direction.

\section{Vortex surfaces and Dirac eigenmode densities} 
\begin{figure*}[h]
  \psfrag{vortex correlation}[0][-1][1][0]{\small vortex correlation}
  \psfrag{number of attached plaquettes}[-1][0][1][0]{\small number of
  attached plaquettes} \psfrag{1st mode}[-1][0][.8][0]{\footnotesize
  $1^{st}$ mode} \psfrag{20th mode}[-1][0][.8][0]{\footnotesize $20^{th}$
  mode} \begin{tabular}{cc}
    \includegraphics[scale=0.55]{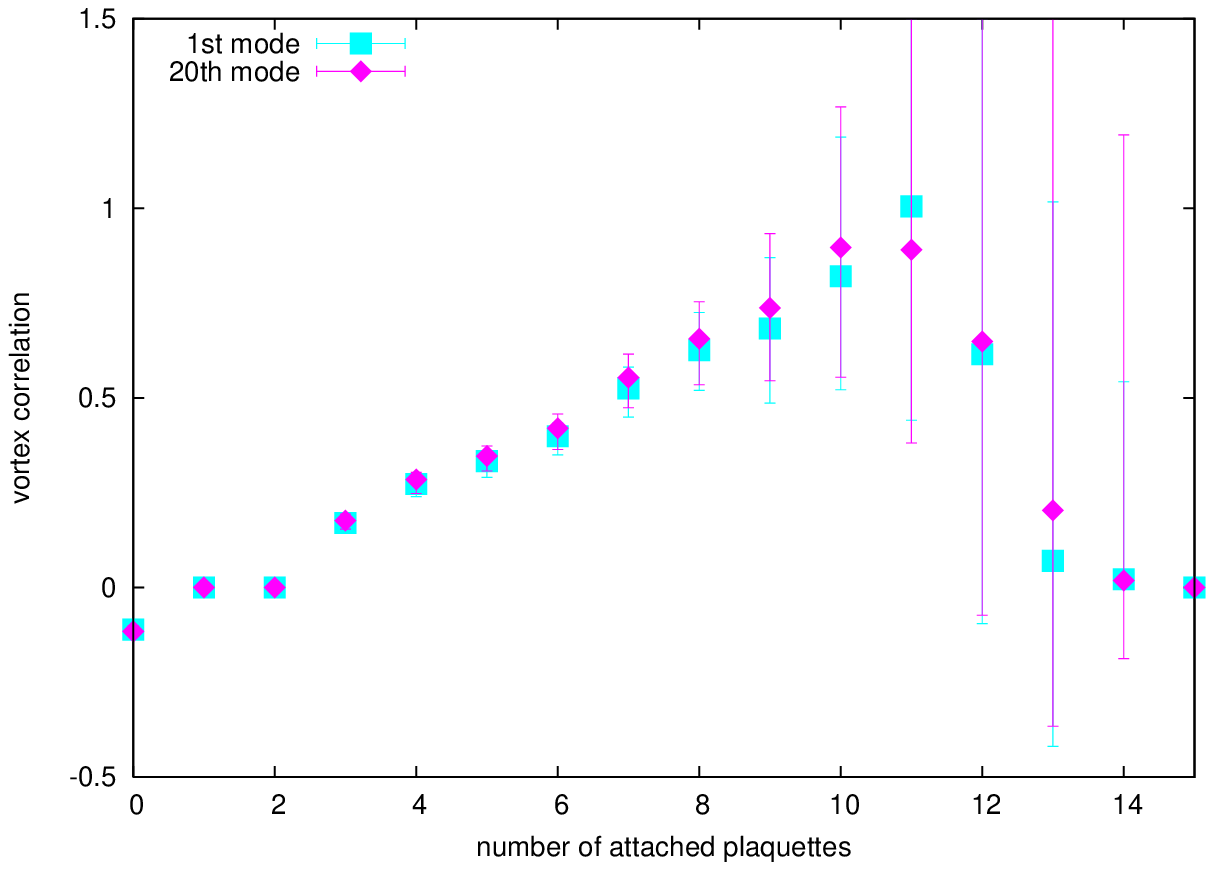}&
    \includegraphics[scale=0.55]{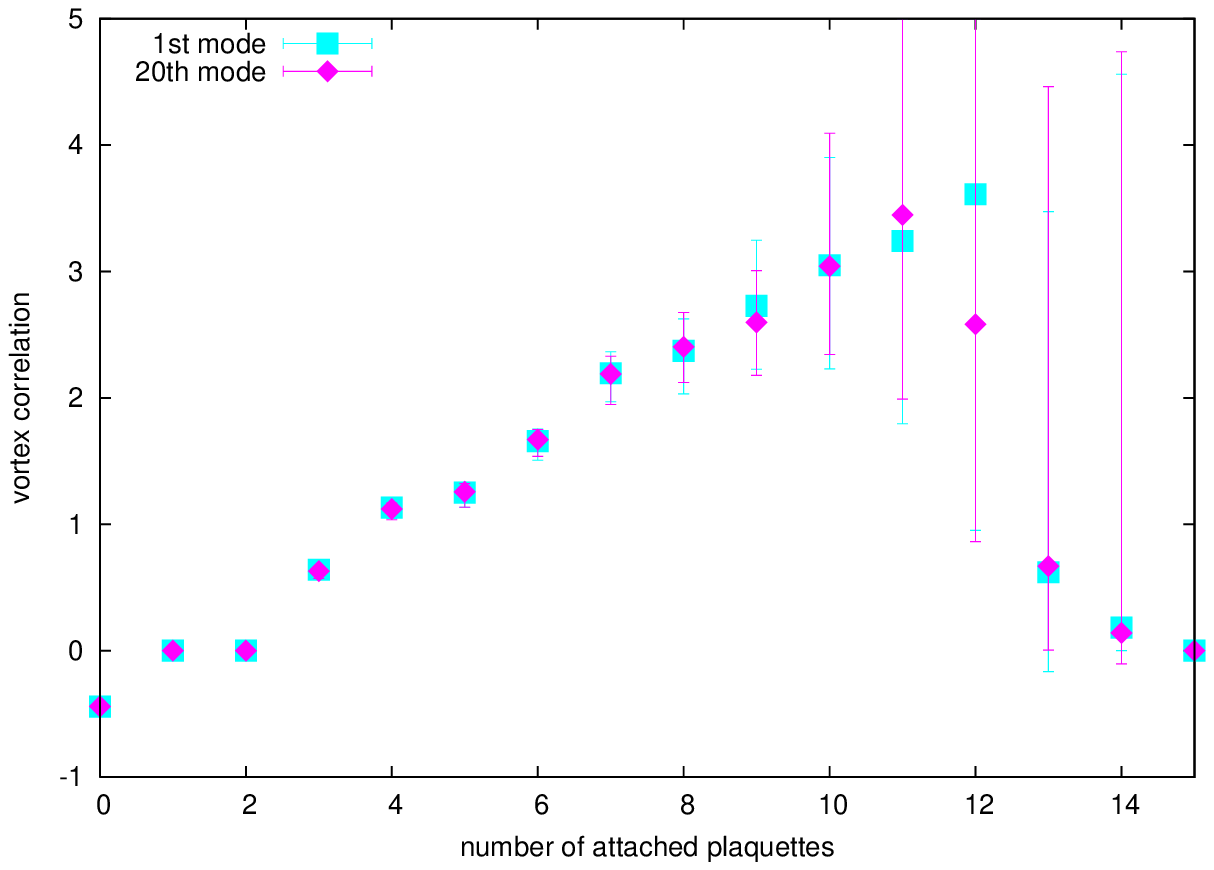}
  \end{tabular} \caption{Vortex correlation $C_\l(N_{\rm v})$ for asqtad
  staggered eigenmodes on a $20^4$ lattice at $\b_{LW}=3.3$, full (left)
  and center-projected (right) configurations.}
\label{vcasqtad} 
\end{figure*}

In order to
clarify the role of the vortices in the topological structure of the
vacuum, we study the correlation between the density $\rho_\lambda(x)$
of an eigenmode with eigenvalue $\lambda$, and the number of vortex
plaquettes (identified via center projection) which meet at a
site. This is relevant because of the picture advanced by Engelhardt
and Reinhardt \cite{ER}, in which topological charge is associated with
vortex intersections ($N_{\rm v}=8$) and ``writhings" ($N_{\rm v}=6$).
Following Kovalenko et al.\ \cite{ITEP}, we define the correlator
\begin{equation}
C_\lambda(N_{\rm v}) = \frac{\sum_{p_i}\sum_{x\in H}(V\rho_\lambda(x)-1)}
{\sum_{p_i}\sum_{x\in H}1} ~.
\label{equ:correlator}
\end{equation}
Here the sum is over sites $p_i$ on the dual lattice which belong to $N_v$
plaquettes on the vortex surface (as identified from center projection);
$V$ is the lattice volume.  At each such vortex site on the dual lattice
there is a second sum ($x\in H$) over sites in a hypercube on the original
lattice surrounding $p_i$.  In Fig.\ \ref{vcasqtad} we display the data
for $C_\l(N_{\rm v})$ vs.\ $N_{\rm v}$ computed for eigenmodes of the
asqtad Dirac operator in the full and center-projected configurations. We
find that the values of $C_\l(N_{\rm v})$ obtained from eigenmodes
in the full configurations are only about a factor of four smaller
than the corresponding values in the center-projected configurations,
and the figures look much the same. The most important feature, in
our opinion, is the fact that the correlator increases steadily with
increasing number of the vortex plaquettes $N_{\rm v}$.  The eigenmode
density seems to be significantly enhanced at vortex sites with large
$N_{\rm v}$.  Our results for eigenmodes of the overlap operator are
similar, and consistent with the results reported by Kovalenko et al.\
in Ref.\ \cite{ITEP}.

\begin{figure*}[p]
  \begin{tabular}{ccc}
    \includegraphics[scale=0.33]{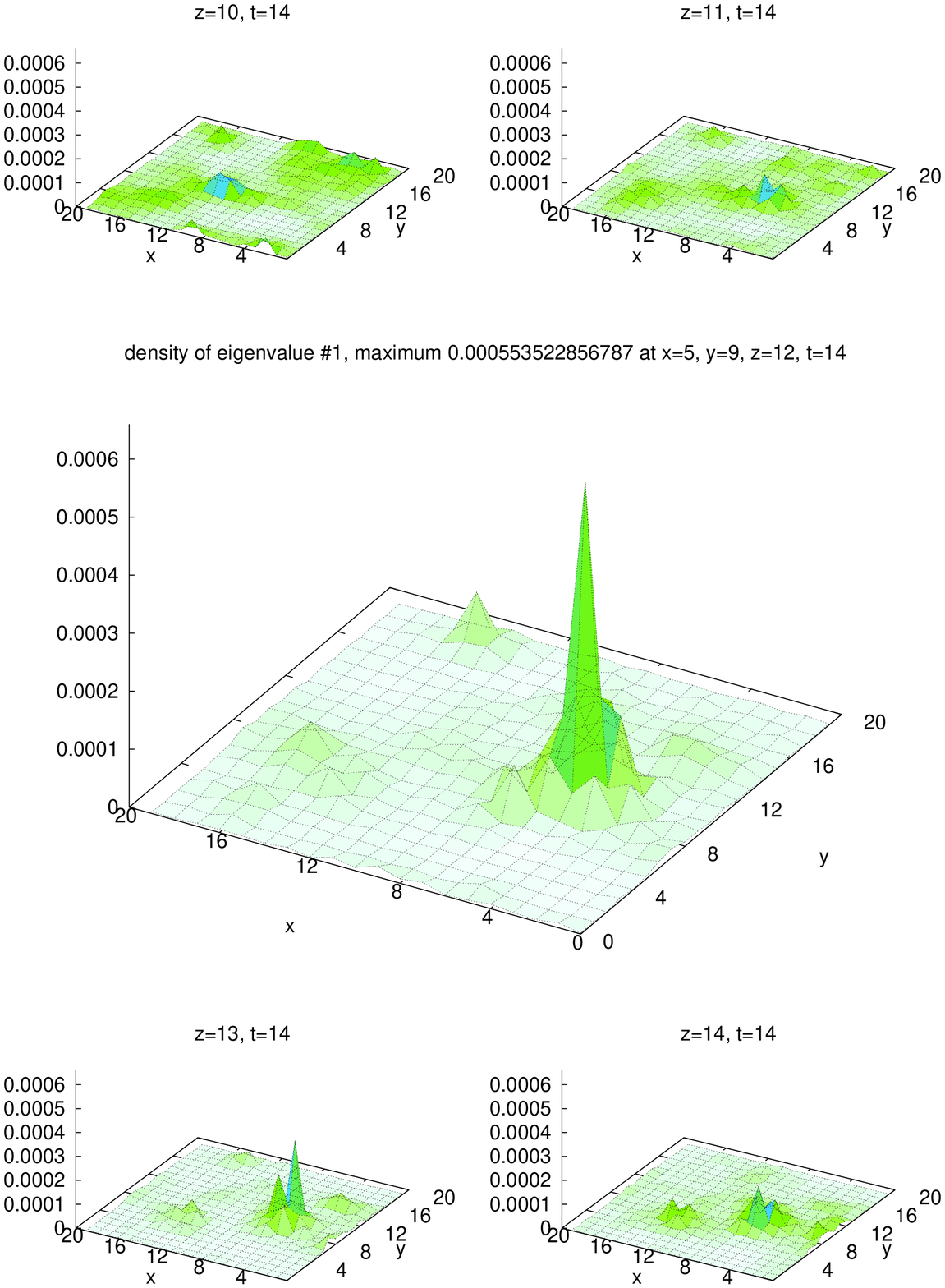}&$\;$&
    \includegraphics[scale=0.33]{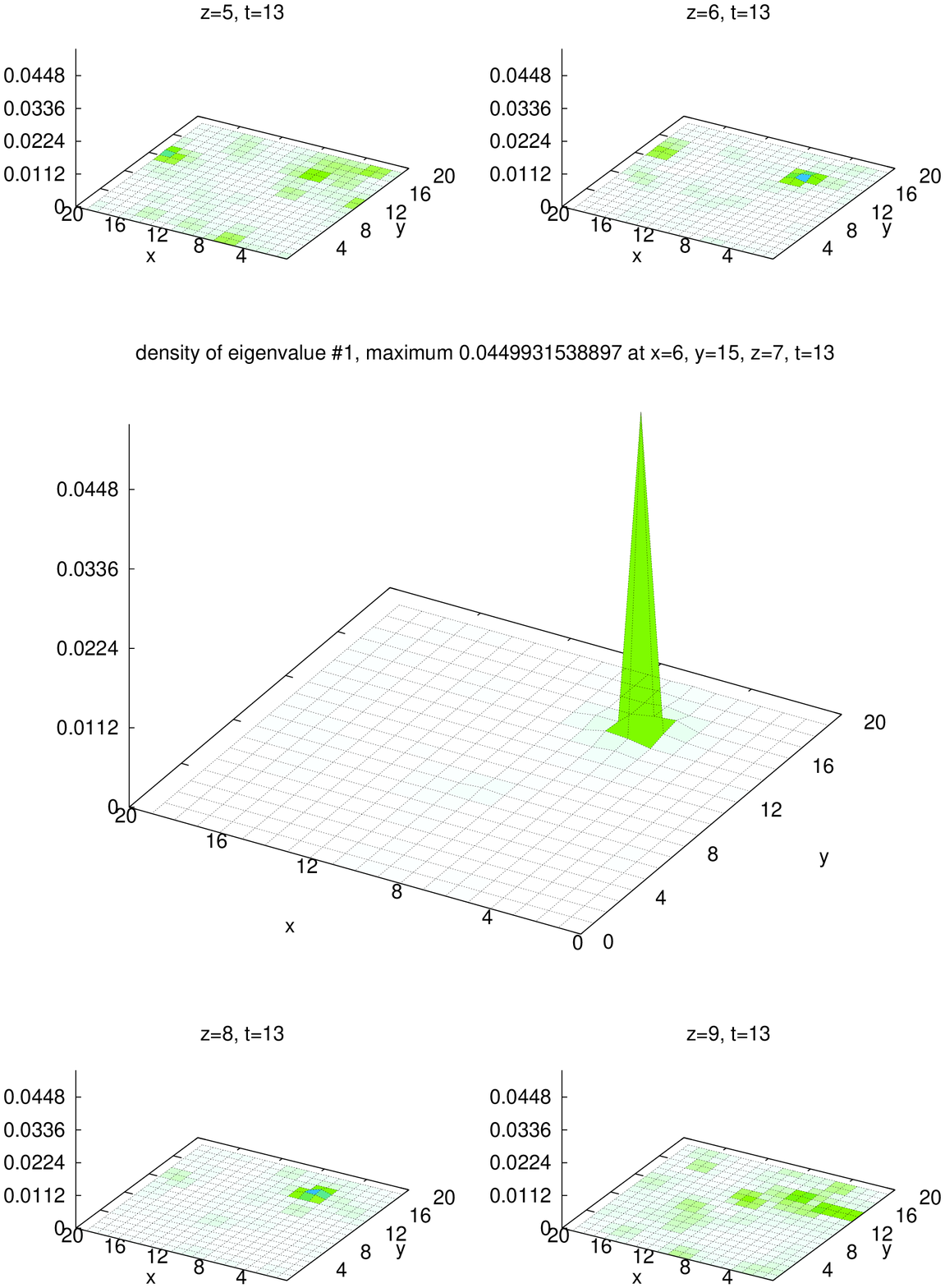}
  \end{tabular}
  \caption{Maximum density peak (center) of the first \textbf{asqtad}
  eigenmode on a $20^4$-lattice at $\b_{LW}=3.3$ with upper (above) and
  lower (below) z-slices of the same t-slice. Eigenmodes are computed on
  (full (left) and center-projected (right) lattices (notice different
  scales!).}
  \label{asqtad_peak} 
\end{figure*}

\begin{figure*}[p]
  \begin{tabular}{ccc}
    \includegraphics[scale=0.33]{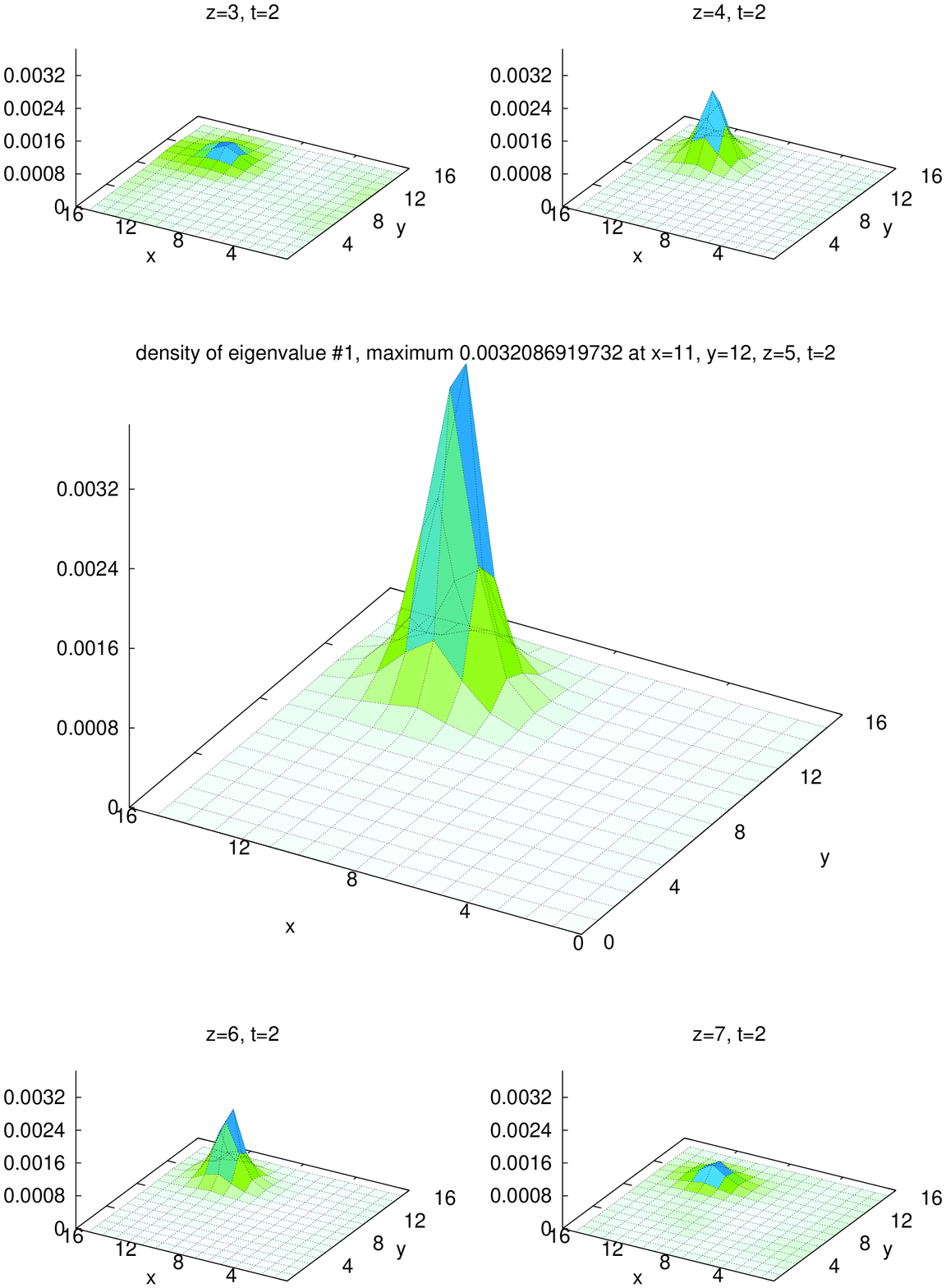}&$\;$&
    \includegraphics[scale=0.33]{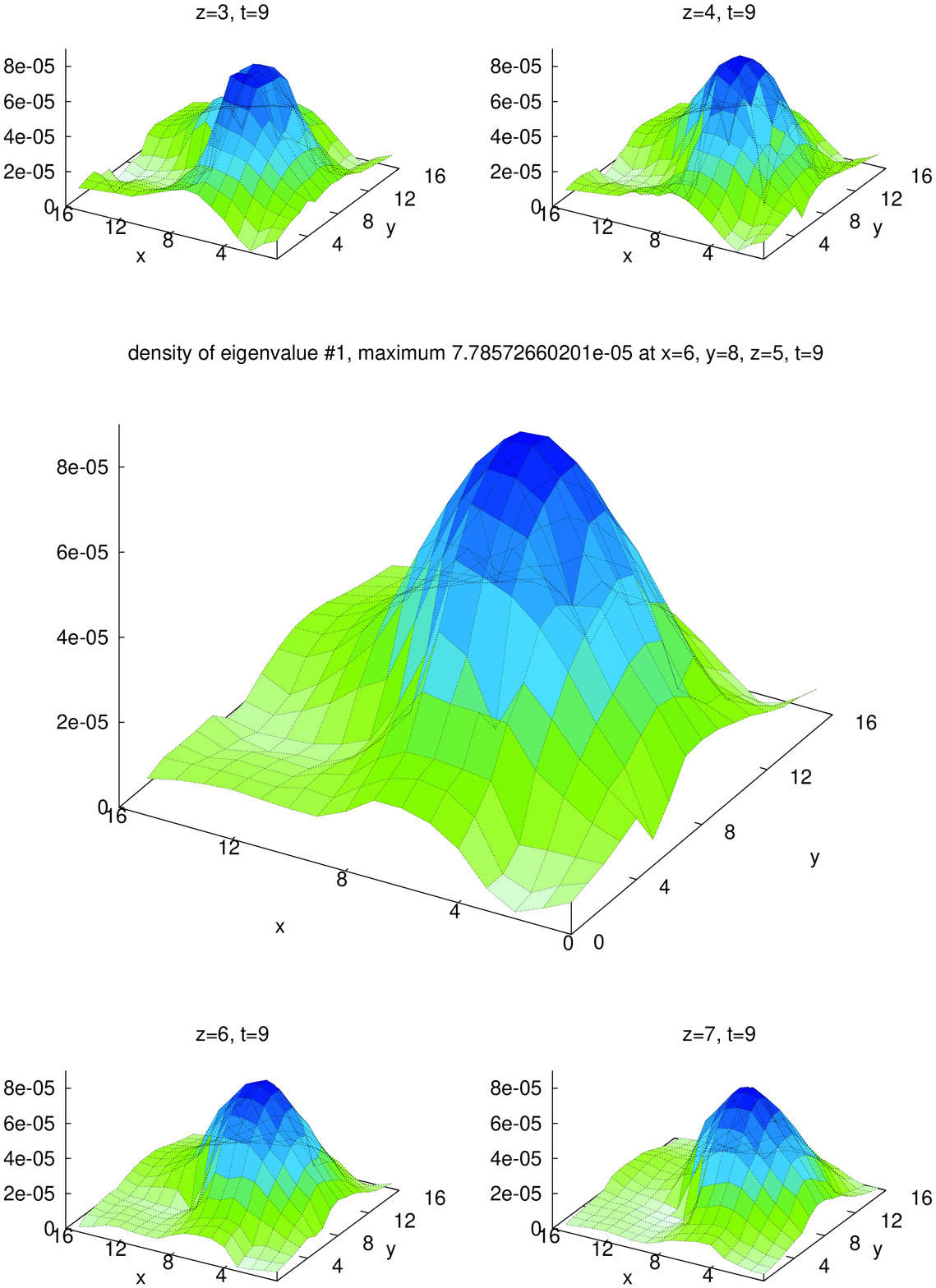}
  \end{tabular}
  \caption{Maximum density peak (center) of the first \textbf{overlap}
  eigenmode on a $16^4$-lattice at $\b_{LW}=3.3$ with upper (above) and
  lower (below) z-slices of the same t-slice. Eigenmodes are computed on
  (a) full lattices, and (b) center-projected lattices (notice different
  scales!).}
  \label{olap_peak} 
\end{figure*}

The correlations shown provide some degree of evidence that low-lying
Dirac eigenmodes concentrate preferentially at regions on the center
vortex surface where there are self-intersections or ``writhing''-points,
in agreement with the general picture advanced by Engelhardt and
Reinhardt \cite{ER}. It is then natural to ask whether there is any
supporting evidence that the eigenmode density is especially concentrated
in point-like regions. To check this, we simply inspect sample plots of
$\rho_\l(x)$ throughout the lattice volume.  In Fig.\ \ref{asqtad_peak} we
display our data for the lowest eigenmode of the asqtad Dirac operator, in
some two-dimensional slices of the four-dimensional lattice volume taken
in the neighborhood of the point where $\rho_\l(x)$ is largest. Each
lattice, unprojected (left) and center-projected (right), contains
several sharp peaks of this kind; the concentration of eigenmode density
is in a point-like region, rather than being spread over a submanifold of
higher dimensionality. Figure \ref{olap_peak} shows the same type of data
for a zero mode of the overlap Dirac operator on $16^4$ lattices. For
full configurations the eigenmode density again is concentrated in
a point-like region.  For the overlap we have already noted that the
spectrum evaluated in center-projected configurations is unrelated to \cb,
and indeed, instead of having a sharp peak, the eigenmode concentration
in this case extends over most of the lattice volume.

\section{Conclusions}

We find that the thin vortices found in center projection give rise to
a low-lying spectrum of near-zero Dirac eigenmodes, provided that the
chiral symmetry of the Dirac operator does not depend on the smoothness
of the lattice configuration. Thus, the vortex excitations of the
vortex-only lattice carry not only the information about confinement,
but are also responsible for \cb ~via the Banks-Casher relation. There
are significant correlations between center vortices and the low-lying
modes of both the asqtad and overlap Dirac operators. These eigenmodes
have their largest concentrations in point-like regions, rather than
on submanifolds of higher dimensionality. Taken together, correlations
and dimensionality support the picture of a center vortex origin for
topological charge, and indicate that center vortices have a strong effect
on the properties of low-lying eigenmodes of the Dirac operator. A more
detailed presentation is found in Ref.\ \cite{Hollwieser:2008tq}.

\end{document}